\documentclass[10pt,a4paper,onecolumn]{article}
\usepackage{marginnote}
\usepackage{graphicx}
\usepackage{xcolor}
\usepackage{authblk,etoolbox}
\usepackage{titlesec}
\usepackage{calc}
\usepackage{tikz}
\usepackage{hyperref}
\hypersetup{colorlinks,breaklinks=true,
            urlcolor=[rgb]{0.0, 0.5, 1.0},
            linkcolor=[rgb]{0.0, 0.5, 1.0}}
\usepackage{caption}
\usepackage{tcolorbox}
\usepackage{amssymb,amsmath}
\usepackage{ifxetex,ifluatex}
\usepackage{seqsplit}
\usepackage{xstring}

\newcommand{\pmwd}{{\usefont{T1}{nova}{m}{sl}pmwd}}

\usepackage{float}
\let\origfigure\figure
\let\endorigfigure\endfigure
\renewenvironment{figure}[1][2] {
    \expandafter\origfigure\expandafter[H]
} {
    \endorigfigure
}

\usepackage{fixltx2e} 
\usepackage[
  backend=biber,
]{biblatex}
\bibliography{pmwd.bib}

\let\textttOrig=\texttt
\def\texttt#1{\expandafter\textttOrig{\seqsplit{#1}}}
\renewcommand{\seqinsert}{\ifmmode
  \allowbreak
  \else\penalty6000\hspace{0pt plus 0.02em}\fi}

\makeatletter
\let\href@Orig=\href
\def\href@Urllike#1#2{\href@Orig{#1}{\begingroup
    \def\Url@String{#2}\Url@FormatString
    \endgroup}}
\def\href@Notdoi#1#2{\def\tempa{#1}\def\tempb{#2}%
  \ifx\tempa\tempb\relax\href@Urllike{#1}{#2}\else
  \href@Orig{#1}{#2}\fi}
\def\href#1#2{%
  \IfBeginWith{#1}{https://doi.org}%
  {\href@Urllike{#1}{#2}}{\href@Notdoi{#1}{#2}}}
\makeatother

\newlength{\cslhangindent}
\newlength{\csllabelwidth}
\newenvironment{CSLReferences}[2] 
 {
  \setlength{\parindent}{0pt}
  \ifodd #1 \everypar{\setlength{\hangindent}{\cslhangindent}}\ignorespaces\fi
  \ifnum #2 > 0
  \setlength{\parskip}{#2\baselineskip}
  \fi
 }%
 {}
\usepackage{calc}

\usepackage[top=3.5cm, bottom=3cm, right=1.5cm, left=1.0cm,
            headheight=2.2cm, reversemp, includemp, marginparwidth=4.5cm]{geometry}

\titleformat{\section}
  {\normalfont\sffamily\Large\bfseries}
  {}{0pt}{}
\titleformat{\subsection}
  {\normalfont\sffamily\large\bfseries}
  {}{0pt}{}
\titleformat{\subsubsection}
  {\normalfont\sffamily\bfseries}
  {}{0pt}{}
\titleformat*{\paragraph}
  {\sffamily\normalsize}

\usepackage{fancyhdr}
\makeatletter
\let\ps@plain\ps@fancy
\fancyheadoffset[L]{4.5cm}
\fancyfootoffset[L]{4.5cm}

\definecolor{linky}{rgb}{0.0, 0.5, 1.0}

\newtcolorbox{repobox}
   {colback=red, colframe=red!75!black,
     boxrule=0.5pt, arc=2pt, left=6pt, right=6pt, top=3pt, bottom=3pt}

\newcommand{\ExternalLink}{%
   \tikz[x=1.2ex, y=1.2ex, baseline=-0.05ex]{%
       \begin{scope}[x=1ex, y=1ex]
           \clip (-0.1,-0.1)
               --++ (-0, 1.2)
               --++ (0.6, 0)
               --++ (0, -0.6)
               --++ (0.6, 0)
               --++ (0, -1);
           \path[draw,
               line width = 0.5,
               rounded corners=0.5]
               (0,0) rectangle (1,1);
       \end{scope}
       \path[draw, line width = 0.5] (0.5, 0.5)
           -- (1, 1);
       \path[draw, line width = 0.5] (0.6, 1)
           -- (1, 1) -- (1, 0.6);
       }
   }

\patchcmd{\@maketitle}{center}{flushleft}{}{}
\patchcmd{\@maketitle}{center}{flushleft}{}{}
\patchcmd{\@maketitle}{\LARGE}{\LARGE\sffamily}{}{}
\def\maketitle{{%
  
  \AB@maketitle}}
\makeatletter
\renewcommand\AB@affilsepx{ \protect\Affilfont}
\renewcommand\AB@affilnote[1]{{\bfseries #1}\hspace{3pt}}
\renewcommand{\affil}[2][]%
   {\newaffiltrue\let\AB@blk@and\AB@pand
      \if\relax#1\relax\def\AB@note{\AB@thenote}\else\def\AB@note{#1}%
        \setcounter{Maxaffil}{0}\fi
        \begingroup
        \let\href=\href@Orig
        \let\texttt=\textttOrig
        \let\protect\@unexpandable@protect
        \def\thanks{\protect\thanks}\def\footnote{\protect\footnote}%
        \@temptokena=\expandafter{\AB@authors}%
        {\def\\{\protect\\\protect\Affilfont}\xdef\AB@temp{#2}}%
         \xdef\AB@authors{\the\@temptokena\AB@las\AB@au@str
         \protect\\[\affilsep]\protect\Affilfont\AB@temp}%
         \gdef\AB@las{}\gdef\AB@au@str{}%
        {\def\\{, \ignorespaces}\xdef\AB@temp{#2}}%
        \@temptokena=\expandafter{\AB@affillist}%
        \xdef\AB@affillist{\the\@temptokena \AB@affilsep
          \AB@affilnote{\AB@note}\protect\Affilfont\AB@temp}%
      \endgroup
       \let\AB@affilsep\AB@affilsepx
}
\makeatother

\renewcommand\Affilfont{\sffamily\small\mdseries}
\ifnum 0\ifxetex 1\fi\ifluatex 1\fi=0 
  \usepackage[T1]{fontenc}
  \usepackage[utf8]{inputenc}

\else 
  \ifxetex
    \usepackage{mathspec}
    \usepackage{fontspec}

  \else
    \usepackage{fontspec}
  \fi
  \defaultfontfeatures{Ligatures=TeX,Scale=MatchLowercase}

\fi
\IfFileExists{upquote.sty}{\usepackage{upquote}}{}
\IfFileExists{microtype.sty}{%
\usepackage{microtype}
\UseMicrotypeSet[protrusion]{basicmath} 
}{}

\usepackage{hyperref}
\hypersetup{unicode=true,
            pdftitle={: A Differentiable Cosmological Particle-Mesh N-body Library},
            pdfborder={0 0 0},
            breaklinks=true}
\usepackage{longtable,booktabs}

\let\addcontentslineOrig=\addcontentsline
\def\addcontentsline#1#2#3{\bgroup
  \let\texttt=\textttOrig\addcontentslineOrig{#1}{#2}{#3}\egroup}
\let\markbothOrig\markboth
\def\markboth#1#2{\bgroup
  \let\texttt=\textttOrig\markbothOrig{#1}{#2}\egroup}
\let\markrightOrig\markright
\def\markright#1{\bgroup
  \let\texttt=\textttOrig\markrightOrig{#1}\egroup}

\usepackage{graphicx,grffile}
\makeatletter
\def\maxwidth{\ifdim\Gin@nat@width>\linewidth\linewidth\else\Gin@nat@width\fi}
\def\maxheight{\ifdim\Gin@nat@height>\textheight\textheight\else\Gin@nat@height\fi}
\makeatother
\setkeys{Gin}{width=\maxwidth,height=\maxheight,keepaspectratio}
\IfFileExists{parskip.sty}{%
\usepackage{parskip}
}{
\setlength{\parindent}{0pt}
\setlength{\parskip}{6pt plus 2pt minus 1pt}
}
\ifx\paragraph\undefined\else
\let\oldparagraph\paragraph
\renewcommand{\paragraph}[1]{\oldparagraph{#1}\mbox{}}
\fi
\ifx\subparagraph\undefined\else
\let\oldsubparagraph\subparagraph
\renewcommand{\subparagraph}[1]{\oldsubparagraph{#1}\mbox{}}
\fi

\title{\pmwd: A Differentiable Cosmological Particle-Mesh \(N\)-body
Library}

        \author[1, 2, 3]{Yin Li}
          \author[2]{Libin Lu}
          \author[2, 3]{Chirag Modi}
          \author[4]{Drew Jamieson}
          \author[1, 5]{Yucheng Zhang}
          \author[6]{Yu Feng}
          \author[2, 7]{Wenda Zhou}
          \author[8, 3]{Ngai Pok Kwan}
          \author[9]{François Lanusse}
          \author[1, 10]{Leslie Greengard}
    
      \affil[1]{Department of Mathematics and Theory, Peng Cheng
Laboratory, Shenzhen, Guangdong 518066, China}
      \affil[2]{Center for Computational Mathematics, Flatiron
Institute, New York, New York 10010, USA}
      \affil[3]{Center for Computational Astrophysics, Flatiron
Institute, New York, New York 10010, USA}
      \affil[4]{Max Planck Institute for Astrophysics, 85748 Garching
bei München, Germany}
      \affil[5]{Department of Physics, New York University, New York,
New York 10003, USA}
      \affil[6]{Berkeley Center for Cosmological Physics, University of
California, Berkeley, California 94720, USA}
      \affil[7]{Center for Data Science, New York University, New York,
New York 10011, USA}
      \affil[8]{Department of Physics, The Chinese University of Hong
Kong, Hong Kong}
      \affil[9]{AIM, CEA, CNRS, Université Paris-Saclay, Université
Paris Diderot, Sorbonne Paris Cité, F-91191 Gif-sur-Yvette, France}
      \affil[10]{Courant Institute, New York University, New York, New
York 10012, USA}
  \date{\vspace{-7ex}}

\begin{document}
\maketitle

\marginpar{

  \begin{flushleft}
  \sffamily\small

  {\bfseries DOI:} \href{https://doi.org/10.21105/joss.XXXXX}{\color{linky}{10.21105/joss.XXXXX}}

  \vspace{2mm}

  {\bfseries Software}
  \begin{itemize}
    \setlength\itemsep{0em}
    \item \href{https://github.com/openjournals/joss-reviews/issues/XXXX}{\color{linky}{Review}} \ExternalLink
    \item \href{https://github.com/eelregit/pmwd}{\color{linky}{Repository}} \ExternalLink
    \item \href{https://doi.org/10.5281/zenodo.XXXXXXX}{\color{linky}{Archive}} \ExternalLink
  \end{itemize}

  \vspace{2mm}

  \par\noindent\hrulefill\par

  \vspace{2mm}

  {\bfseries Editor:} \href{https://editor\_url.xyz}{Editor
Name} \ExternalLink \\
  \vspace{1mm}
    {\bfseries Reviewers:}
  \begin{itemize}
  \setlength\itemsep{0em}
    \item \href{https://github.com/Reviewer A}{@Reviewer A}
    \item \href{https://github.com/Reviewer B}{@Reviewer B}
    \end{itemize}
    \vspace{2mm}

  {\bfseries Submitted:} 2022-XX-XX\\
  {\bfseries Published:} 2023-XX-XX

  \vspace{2mm}
  {\bfseries License}\\
  Authors of papers retain copyright and release the work under a Creative Commons Attribution 4.0 International License (\href{http://creativecommons.org/licenses/by/4.0/}{\color{linky}{CC BY 4.0}}).

    \vspace{4mm}
  {\bfseries In partnership with}\\
  \vspace{2mm}
  \includegraphics[width=4cm]{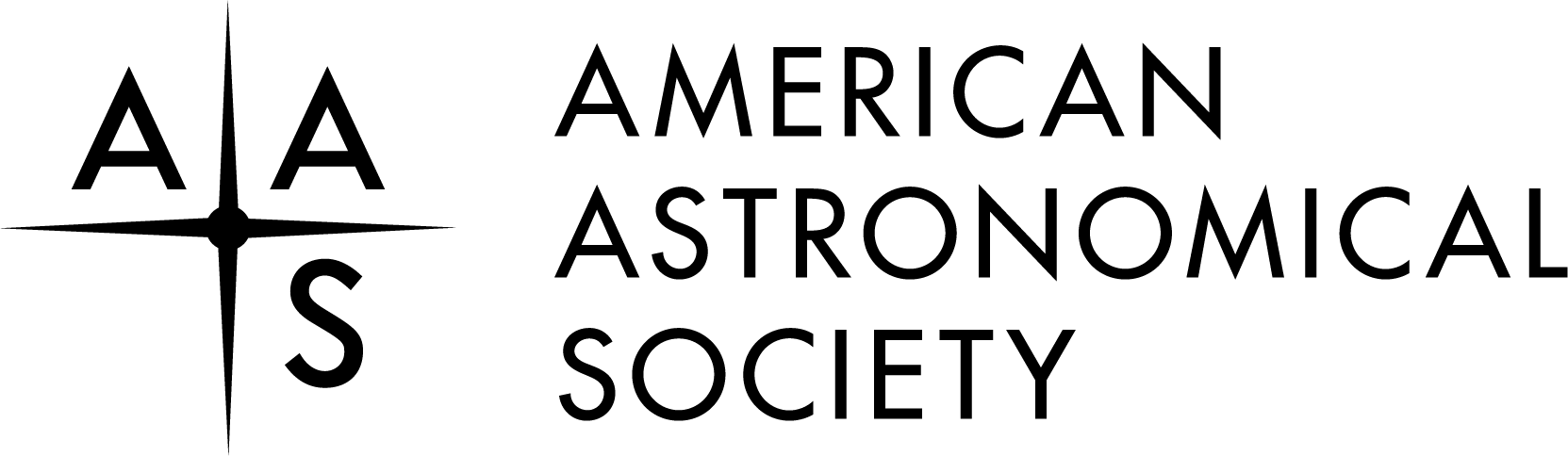}
  \vspace{2mm}
  \newline
  This article and software are linked with research article DOI \href{https://doi.org/10.3847/XXXXX}{\color{linky}{10.3847/XXXXX}}, published in the to
be submitted to Astrophysical Journal Supplement Series.
  
  \end{flushleft}
}

\vspace{2em}

\hypertarget{summary}{%
\section{Summary}\label{summary}}

The formation of the large-scale structure, the evolution and
distribution of galaxies, quasars, and dark matter on cosmological
scales, requires numerical simulations. Differentiable simulations
provide gradients of the cosmological parameters, that can accelerate
the extraction of physical information from statistical analyses of
observational data. The deep learning revolution has brought not only
myriad powerful neural networks, but also breakthroughs including
automatic differentiation (AD) tools and computational accelerators like
GPUs, facilitating forward modeling of the Universe with differentiable
simulations. Because AD needs to save the whole forward evolution
history to backpropagate gradients, current differentiable cosmological
simulations are limited by memory. Using the adjoint method, with
reverse time integration to reconstruct the evolution history, we
develop a differentiable cosmological particle-mesh (PM) simulation
library \pmwd{} (particle-mesh with derivatives) with a low memory cost.
Based on the powerful AD library \texttt{JAX}, \pmwd{} is fully
differentiable, and is highly performant on GPUs.

\vspace{2em}

\begin{figure}
\centering
\includegraphics[width=0.6\textwidth,height=\textheight]{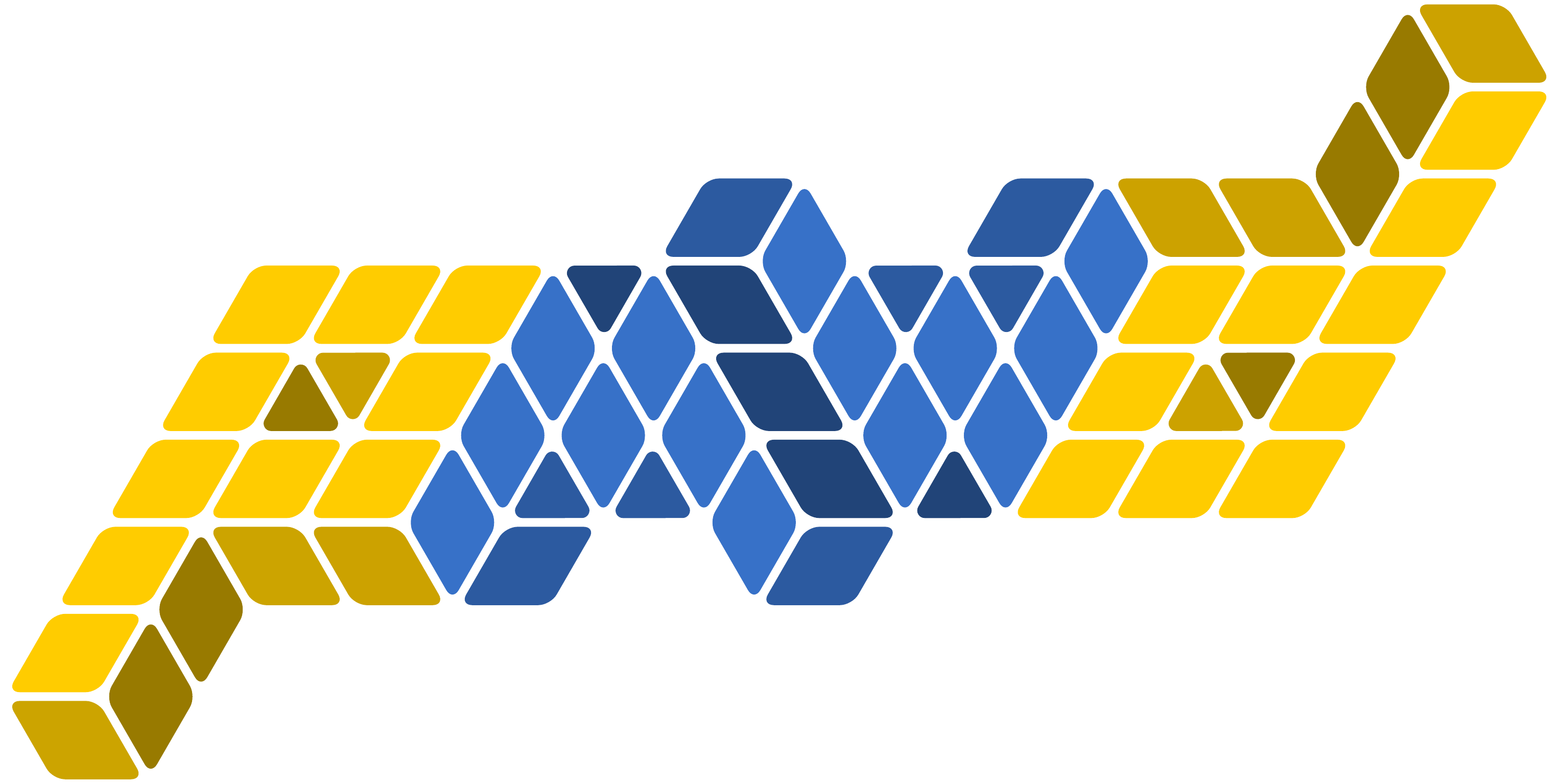}
\caption{\pmwd{} logo. The C2 symmetry of the name symbolizes the
reversibility of the model, which helps to dramatically reduce the
memory cost together with the adjoint method. \label{fig:logo}}
\end{figure}

\clearpage

\hypertarget{statement-of-need}{%
\section{Statement of Need}\label{statement-of-need}}

Current established workflows of statistical inference from cosmological
datasets involve reducing cleaned data to summary statistics like the
power spectrum, and predicting these statistics using perturbation
theories, semi-analytic models, or simulation-calibrated emulators.
Rapid advances in accelerator technology like GPUs opens the possibility
of direct simulation-based forward modeling and inference (Cranmer,
Brehmer, and Louppe 2020), even at level of the fields before their
compression into summary statistics. The forward modeling approach
naturally account for the cross-correlation of different observables,
and can easily incorporate systematic errors. In addition, model
differentiability can accelerate parameter constraint with
gradient-based optimization and inference. A differentiable field-level
forward model combines the two features and is able to constrain
physical parameters together with the initial conditions of the
Universe.

The first differentiable cosmological simulations, such as BORG and
ELUCID (Jasche and Wandelt 2013; Wang et al. 2014), were developed
before the advent of modern AD systems, and were based on
implementations of analytic derivatives. Later codes including
\texttt{FastPM} and \texttt{FlowPM} (Seljak et al. 2017; Modi, Lanusse,
and Seljak 2021) compute gradients using AD engines, namely
\texttt{vmad} (written by the same authors) and \texttt{TensorFlow},
respectively. Both analytic differentiation and AD backpropagate the
gradients through the whole history, thus requires saving the states at
all time steps in memory. Therefore, they are subject to a trade-off
between time and space/mass resolution, and typically can integrate for
only tens of time steps, unlike the standard non-differentiable
simulations.

Alternatively, the adjoint method provides systematic ways of deriving
model gradients under constraints (Pontryagin 1962), such as the
\(N\)-body equations of motion in the simulated Universe (Li et al.
2022). The adjoint method evolves a dual set of equations backward in
time, dependent on the states in the forward run, which can be recovered
by reverse time integration for reversible dynamics, thereby
dramatically reducing the memory cost (Chen et al. 2018). Our logo in
\autoref{fig:logo} is inspired by such reversibility as well as the
\texttt{JAX} artistic style. Furthermore, we take the
discretize-then-optimize approach (e.g., Gholami, Keutzer, and Biros
2019) to ensure gradients propagate backward along the same discrete
trajectory as taken by the forward time integration. We derive and
validate our adjoint method in Li et al. (2022) in more details. The
table below compares the differentiable cosmological simulation codes.

Being both computation and memory efficient, \pmwd{} enables larger and
more accurate forward modeling, thus will improve gradient based
optimization and inference. Differentiable analytic, semi-analytic, and
deep learning components can run based on or in parallel with \pmwd{}
simulations. Examples include a growth function emulator (Kwan et al.
2022) and our ongoing work on spatiotemporal optimization of the PM
gravity solver (Zhang et al. in prep). In the future, \pmwd{} will also
facilitate the modeling of various cosmological observables and the
understanding of the astrophysics at play.

\begin{longtable}[]{@{}rcccc@{}}
\toprule()
code & OSS & gradient & mem efficient & hardware \\
\midrule()
\endhead
BORG & & analytic & & CPU \\
ELUCID & & analytic & & CPU \\
\texttt{FastPM-vmad} & \(\checkmark\) & AD & & CPU \\
\texttt{FlowPM} & \(\checkmark\) & AD & & GPU/CPU \\
\pmwd{} & \(\checkmark\) & adjoint & \(\checkmark\) & GPU/CPU \\
\bottomrule()
\end{longtable}

\hypertarget{acknowledgements}{%
\section{Acknowledgements}\label{acknowledgements}}

The Flatiron Institute is supported by the Simons Foundation.

\hypertarget{references}{%
\section*{References}\label{references}}
\addcontentsline{toc}{section}{References}

\hypertarget{refs}{}
\begin{CSLReferences}{1}{0}
\leavevmode\vadjust pre{\hypertarget{ref-NeuralODE}{}}%
Chen, Ricky T. Q., Yulia Rubanova, Jesse Bettencourt, and David K
Duvenaud. 2018. {``Neural {Ordinary Differential Equations}.''} In
\emph{Advances in {Neural Information Processing Systems} 31}, edited by
S. Bengio, H. Wallach, H. Larochelle, K. Grauman, N. Cesa-Bianchi, and
R. Garnett, 6571--83. {Curran Associates, Inc.}
\url{http://papers.nips.cc/paper/7892-neural-ordinary-differential-equations.pdf}.

\leavevmode\vadjust pre{\hypertarget{ref-CranmerEtAl2020}{}}%
Cranmer, Kyle, Johann Brehmer, and Gilles Louppe. 2020. {``The Frontier
of Simulation-Based Inference.''} \emph{Proceedings of the National
Academy of Sciences} 117 (48): 30055--62.
\url{https://doi.org/10.1073/pnas.1912789117}.

\leavevmode\vadjust pre{\hypertarget{ref-ANODE}{}}%
Gholami, Amir, Kurt Keutzer, and George Biros. 2019. {``ANODE:
Unconditionally Accurate Memory-Efficient Gradients for Neural ODEs.''}
\emph{arXiv Preprint arXiv:1902.10298}.

\leavevmode\vadjust pre{\hypertarget{ref-BORG}{}}%
Jasche, Jens, and Benjamin D. Wandelt. 2013. {``Bayesian Physical
Reconstruction of Initial Conditions from Large-Scale Structure
Surveys.''} \emph{Monthly Notices of the Royal Astronomical Society} 432
(2): 894--913. \url{https://doi.org/10.1093/mnras/stt449}.

\leavevmode\vadjust pre{\hypertarget{ref-KwanModiEtAl2022}{}}%
Kwan, Ngai Pok, Chirag Modi, Yin Li, and Shirley Ho. 2022. {``Emulating
Cosmological Growth Functions with b-Splines.''} In \emph{NeurIPS 2022
Machine Learning and the Physical Sciences}.
\url{http://arxiv.org/abs/2211.06564}.

\leavevmode\vadjust pre{\hypertarget{ref-adjoint}{}}%
Li, Yin, Chirag Modi, Drew Jamieson, Yucheng Zhang, Libin Lu, Yu Feng,
François Lanusse, and Greengard Leslie. 2022. {``Differentiable
Cosmological Simulation with Adjoint Method.''} \emph{To Be Submitted to
Astrophysical Journal Supplement Series}.
\url{https://arxiv.org/abs/2211.THIS-1}.

\leavevmode\vadjust pre{\hypertarget{ref-FlowPM}{}}%
Modi, Chirag, François Lanusse, and Uroš Seljak. 2021. {``{FlowPM}:
Distributed Tensorflow Implementation of the {FastPM} Cosmological
{N-body} Solver.''} \emph{Astronomy and Computing} 37 (October): 100505.
\url{https://doi.org/10.1016/j.ascom.2021.100505}.

\leavevmode\vadjust pre{\hypertarget{ref-Pontryagin1962}{}}%
Pontryagin, Lev Semenovich. 1962. \emph{The Mathematical Theory of
Optimal Processes}. CRC press.

\leavevmode\vadjust pre{\hypertarget{ref-SeljakEtAl2017}{}}%
Seljak, Uroš, Grigor Aslanyan, Yu Feng, and Chirag Modi. 2017.
{``Towards Optimal Extraction of Cosmological Information from Nonlinear
Data.''} \emph{Journal of Cosmology and Astroparticle Physics} 2017
(12): 009--9. \url{https://doi.org/10.1088/1475-7516/2017/12/009}.

\leavevmode\vadjust pre{\hypertarget{ref-ELUCID}{}}%
Wang, Huiyuan, H. J. Mo, Xiaohu Yang, Y. P. Jing, and W. P. Lin. 2014.
{``{ELUCID} - Exploring the Local Universe with the Reconstructed
Initial Density Field. {I}. Hamiltonian Markov Chain Monte Carlo Method
with Particle Mesh Dynamics.''} \emph{The Astrophysical Journal} 794
(1): 94. \url{https://doi.org/10.1088/0004-637X/794/1/94}.

\leavevmode\vadjust pre{\hypertarget{ref-ZhangLiEtAl}{}}%
Zhang, Yucheng, Yin Li, Drew Jamieson, and et al. in prep.
{``Spatiotemporal Optimization of Particle-Mesh Simulations,''} in prep.

\end{CSLReferences}

\end{document}